\documentclass[prl,superscriptaddress,graphicx,reprint]{revtex4-2}
\usepackage{amsmath, bm, xcolor,graphicx,inputenc,textcomp,siunitx}
\usepackage[hyperindex,breaklinks,colorlinks=true, urlcolor=blue, citecolor=blue, anchorcolor=blue, linkcolor=blue]{hyperref}
\usepackage[all]{hypcap}
\usepackage[capitalise]{cleveref}

\newcommand*{\added}{}
\newcommand{\murm}{\si{\micro}}

\usepackage{ulem}[normalem] 

\begin{document}

\title{Magnetic field amplification by a nonlinear electron streaming instability}

\author{J. R. Peterson}
\email[]{jrpete@stanford.edu}
\affiliation{Physics Department, Stanford University, Stanford, CA 94305}
\affiliation{SLAC National Accelerator Laboratory, Menlo Park, CA 94025}
\author{S. Glenzer}
\affiliation{SLAC National Accelerator Laboratory, Menlo Park, CA 94025}
\author{F. Fiuza}
\email[]{fiuza@slac.stanford.edu}
\affiliation{SLAC National Accelerator Laboratory, Menlo Park, CA 94025}

\date{\today}

\begin{abstract}
  Magnetic field amplification by relativistic streaming plasma instabilities is central to a wide variety of high-energy astrophysical environments as well as to laboratory scenarios associated with intense lasers and electron beams. We report on a new secondary nonlinear instability which arises for relativistic dilute electron beams after the saturation of the linear Weibel instability. This instability grows due to the transverse magnetic pressure associated with the beam current filaments, which cannot be quickly neutralized due to the inertia of background ions. We show that it can amplify the magnetic field strength and spatial scale by orders of magnitude, leading to large-scale plasma cavities with strong magnetic field and to very efficient conversion of the beam kinetic energy into magnetic energy. The instability growth rate, saturation level, and scale length are derived analytically and shown to be in good agreement with fully-kinetic simulations.
\end{abstract}


\maketitle

    Relativistic streaming plasma instabilities are ubiquitous in energetic plasma environments. They play an important role in magnetic field amplification \cite{silva03,bell04,broderick12,sironi14}, particle acceleration \cite{spitkovsky08,martins09,sironi15}, and radiation emission \cite{medvedev99,gruzinov99} in astrophysical environments such as those associated with collisionless shocks and relativistic jets. They are also common in laboratory intense laser- and beam-plasma interactions connected to studies of laboratory astrophysics \cite{fiuza12,allen12,fox13,huntington15,fiuza20,shukla20}, electron \cite{silva02,gremillet02} and ion \cite{goede17} transport in high-energy-density and inertial fusion plasmas, and novel schemes for compact, bright gamma-ray sources \cite{benedetti18}.

    \added{Among the different processes that arise in relativistic streaming plasmas, the Weibel (or current filamentation) instability \cite{weibel59,fried59} has garnered significant attention as a leading mechanism for the rapid amplification of magnetic fields \cite{silva03,medvedev05,spitkovsky08,fiuza12,grassi17}. 
    However, in dilute-beam systems (beam density $\ll$ background density) comprising most astrophysical and laboratory environments of interest, the magnetic field strength produced is weak \cite{califano98-sat,califano02,sironi14,benedetti18} and its spatial scale is limited to the order of the plasma skin depth \cite{davidson72}. For highly relativistic particles, this scale is orders of magnitude smaller than the particle gyroradius and it is unclear how these fields could control energetic particle dynamics. Nonlinear filament merging \cite{medvedev05} can slowly increase the magnetic field wavelength, but existing ultrarelativistic particle-in-cell (PIC) simulations show that saturation remains limited to small spatial scales and low magnetizations $\epsilon_B \lesssim 10^{-3}$ \cite{sironi14,benedetti18} (with $\epsilon_B$ the ratio of the magnetic energy density to the beam kinetic energy density). Understanding the long-term nonlinear evolution and true saturation behavior of these systems, both in terms of field strength and coherence length, remains a critical open question.}
    
    In this Letter, we report a new nonlinear streaming instability which arises in relativistic dilute electron beams after saturation of the linear Weibel instability. This instability is driven by the magnetic pressure in the beam current filaments, where the background ions cannot effectively screen the current due to their large inertia. The instability can amplify the magnetic field strength and coherence length by orders of magnitude, generating large-scale plasma cavities and efficiently converting the beam kinetic energy into magnetic energy. Analytic theory of the growth rate, saturation level, and scale length of the instability agree well with fully-kinetic simulations, revealing a robust mechanism for large-scale magnetic field amplification in dilute-beam systems.

    
    To study the nonlinear late-time evolution of streaming instabilities in dilute-beam systems, we performed fully kinetic one-(1D), two-(2D), and three-dimensional (3D) simulations with the relativistic PIC code OSIRIS \cite{fonseca02,fonseca08}. We study a spatially-uniform system containing a cold, dilute, ultrarelativistic electron beam of density $n_{b0}$ propagating in a cold background plasma with a density ratio $\alpha=n_{b0}/n_{e0} \ll 1$ and \added{$Zn_{i0}=n_{e0} + n_{b0}$}, where $n_{e0}$ and $n_{i0}$ are the initial densities of the background electrons and ions \added{and $Z$ the ion charge number}. The beam has Lorentz factor $\gamma_{b0} \gg 1$, and the beam \added{electrons, background electrons, and ions} have initial velocities $v_{b0} \approx c$, $v_{e0} = -\alpha v_{b0}$, \added{and $v_{i0} = 0$} in the $x$-direction such that the system is current neutral. 
    
    A large parameter scan in $\alpha$ and $\gamma_{b0}$ was performed to study how the beam parameters affect the long-term evolution of the system. The simulations \added{covered different geometries and dimensionality, with box sizes of $60,000~c/\omega_p$ in 1D, $1000\times1000~(c/\omega_p)^2$ in 2D $x$-$y$, $4000\times4000~(c/\omega_p)^2$ in 2D $y$-$z$, and $2800\times1400\times1400~(c/\omega_p)^3$ in 3D. (In 1D and 2D $y$-$z$ geometries only the directions transverse to the beam propagation are captured.)
    The cell size used ranged between $\Delta x = \Delta y = \Delta z = 0.25~c/\omega_p$ (2D $x$-$y$) and $1.0~c/\omega_p$ (in all others),} with $c$ the speed of light, $\omega_p = (4 \pi n_{e0} e^2 / m_e)^{1/2}$ the background electron plasma frequency, $m_e$ the electron mass, and $-e$ the electron charge. \added{We model a hydrogen plasma, $m_i/Zm_e=1836$, except where noted.} The time step is chosen according to the Courant–Friedrichs–Lewy condition. We use 20 (1D), 9 (2D), and 8 (3D) particles per cell per species and periodic boundary conditions. We have tested different resolutions and numbers of particles per cell to ensure convergence of the results and have used a third order particle interpolation scheme for improved numerical accuracy.

\begin{figure}
      \includegraphics[width=3.375in]{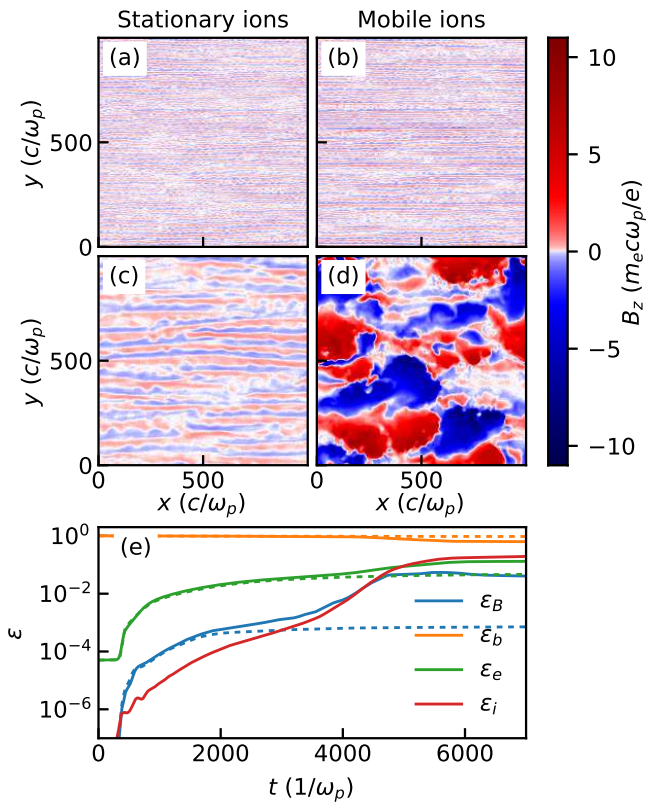}
      \caption{\label{fig:1}
        2D simulation results of a dilute, ultrarelativistic electron beam propagating through an electron-ion plasma. Magnetic field profiles for cases with (a,c) stationary ions and (b,d) mobile ions, taken at (a,b) saturation of the Weibel instability ($t = 1750~\omega_p^{-1}$) and (c,d) saturation of a nonlinear electron streaming instability ($t = 5250~\omega_p^{-1}$). The energy distribution of these simulations is shown in (e) where the solid lines correspond to the mobile ion simulations in (b,d) and dashed lines correspond to stationary ion simulations in (a,c). The beam electron ($\epsilon_b$), background electron ($\epsilon_e$), and ion ($\epsilon_i$) kinetic energy densities as well as the magnetic field energy density ($\epsilon_B$) are all normalized to the initial beam electron kinetic energy density.
      }
    \end{figure}

    We begin by comparing the long-term evolution of the magnetic field with and without mobile ions. The results are illustrated in Fig. \ref{fig:1} for $\alpha = 0.1$ and $\gamma_{b0} = 1000$. \added{We model the 2D $x$-$y$ plane to capture both electrostatic and electromagnetic modes.} The system is initially dominated by the oblique instability \cite{bret10}, which \added{heats the background plasma,} produces very weak magnetic fields, \cite{bret10} and quickly saturates  at $t = 500~\omega_p^{-1}$. \added{Weibel-type filamentary modes then arise with a transverse wavelength $\lambda_B = 2\pi c/\omega_p$, consistent with linear theory \cite{davidson72}, and amplify magnetic fields before saturating at $t = 1750~\omega_p^{-1}$}. Up to this point, the stationary and mobile ion cases are nearly identical as shown in Fig.~\hyperref[fig:1]{\ref*{fig:1}(a,b)}; both simulations show similar filaments and reach a magnetization 
    $\epsilon_B = B^2/(8 \pi n_{b0} \gamma_{b0} m_e c^2) = 4\times10^{-4}$. This low magnetization level is consistent with existing dilute-beam theory; the Weibel modes (based on the magnetic trapping mechanism \cite{davidson72}) and oblique modes \cite{sironi14} saturate with $\epsilon_B \sim \alpha/\gamma_{b0} \sim 10^{-4}$. 
    
    At late times, corresponding to the nonlinear evolution of the system, we see dramatic differences in the magnetic field evolution as illustrated in Fig.~\hyperref[fig:1]{\ref*{fig:1}(c,d)} at $t = 5250~\omega_p^{-1}$. With stationary ions, the filament wavelength slightly increases due to filament merging \cite{medvedev05,gedalin10} but its magnetization is largely unchanged [c.f. Fig.~\hyperref[fig:1]{\ref*{fig:1}(e)}]. In contrast, with mobile ions, the magnetization continues to grow, saturating at nearly two orders of magnitude higher energy and with roughly 20 times larger wavelength $\lambda_B$. Furthermore, the rate of magnetic field growth in this nonlinear phase is exponential as shown in Fig.~\hyperref[fig:1]{\ref*{fig:1}(e)} between $t = 3500~\omega_p^{-1}$ and $t = 5000~\omega_p^{-1}$. To our knowledge this fast exponential amplification of the magnetic field energy and coherence length has not been previously observed and suggests that an important secondary instability dominates the nonlinear evolution of the system.
    
    \begin{figure}
      \includegraphics[width=3.375in]{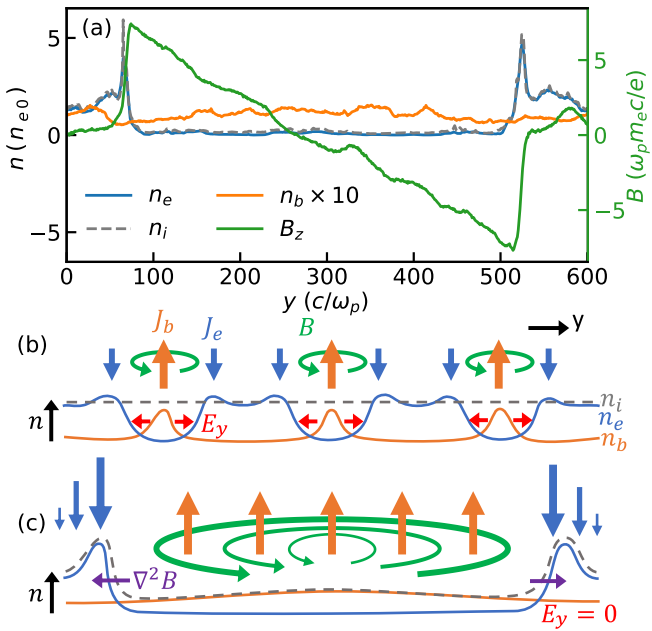}
      \caption{\label{fig:2}
        (a) Cross-sectional profile of the magnetic field and density structure of Fig.~\hyperref[fig:1]{\ref*{fig:1}(d)} taken by averaging the magnetic field and densities field from $x = 390~c/\omega_p$ to $x = 410~c/\omega_p$. The formation of these cavities is sketched in panels (b,c). (b) Fields at saturation of the Weibel instability before ions have moved. (c) Subsequent formation of background plasma cavities expanding due to magnetic pressure.
      }
    \end{figure}
    
    In order to understand the origin of this nonlinear instability 
    \added{we examine the corresponding transverse plasma and magnetic field profile} 
    in Fig.~\hyperref[fig:2]{\ref*{fig:2}(a)}. We observe that the strong magnetic field resides within a plasma cavity 
    \added{with $n_e \approx 0$ 
    and $n_i \approx \alpha n_{e0}$ 
    that charge-neutralizes} the beam. The beam density remains relatively uniform 
    with $n_b \approx n_{b0} = \alpha n_{e0}$. Since the cavity ions are slow to screen the beam current, a strong net current and magnetic field develop in the cavity which are neutralized in the cavity walls by the background electron current. 
    
    The formation of plasma density cavities is a general consequence of strong current filaments in plasmas and has been observed in simulations of a wide variety of systems, including beam-plasma interactions \cite{honda00,sakai02,califano02}, collisionless shocks \cite{fiuza12, ruyer15, naseri18}, and laser-driven ion acceleration \cite{kuznetsov01,bulanov05}. 
    However, previous works have not observed significant growth of the magnetic energy and did not identify a secondary plasma instability related to these structures. We will now \added{discuss how such an instability can arise.}
    
    The first step in cavity formation is saturation of the Weibel instability as illustrated in Fig.~\hyperref[fig:2]{\ref*{fig:2}(b)}. The beam current filaments create magnetic fields which expel the background electrons \added{until a space-charge field $E_x$ arises to arrest the deflection \cite{califano02,bret10}. Eventually, the ions react to $E_x$, exiting the filament and creating a cavity in the background plasma. A few ions remain to charge-neutralize the beam as shown in Fig.~\hyperref[fig:2]{\ref*{fig:2}(c)}, but the beam current is now exposed in the cavity.} The magnetic fields strengthen and the cavity expansion accelerates under increasing magnetic pressure.
    
    The growth rate associated with this instability can be calculated as follows. First, we assume an ultrarelativistic, dilute electron beam ($\gamma_{b0} \gg 1$ and $\alpha \ll 1$) and ion mass $m_i \gg Zm_e$. We consider a cavity containing uniform unscreened beam current $J_b = -e n_{b0} v_{b0}$ \added{in the $x$-direction} and use Amp\`ere's law (neglecting the displacement current) to calculate the magnetic pressure at the cavity walls $P_B = B^2/(8\pi) = \pi J_b^2 \lambda_B^2/(2c^2)$, \added{where $\lambda_B$ is the cavity diameter and dominant wavelength of the magnetic field.}  As the cavity expands, we assume the expelled background plasma accumulates at the wall \added{where $P_B$ exceeds the thermal pressure $P_{th}$. We have verified that $P_B>P_{th}$ locally at the onset of the instability for all of the parameters tested and the cavity can grow even if $\epsilon_B \ll \epsilon_e$ globally as in Fig.~\ref{fig:1}.}
    
    In slab geometry, \added{as in Fig.~\ref{fig:1}, the ion-dominated wall mass is $m_w = m_i n_{e0} A \lambda_B/2Z$ with $A$ its area in the $x$-$z$ plane}. The wall momentum is then $p_w = m_w d(\lambda_B/2)/dt$. The magnetic pressure force will push the wall according to $dp_w/dt = P_B A$ which can be written as
    \begin{equation}
      \frac{d}{d t}\left(\lambda_B \frac{d\lambda_B}{dt}\right) = \frac{\alpha^2 \beta_b^2}{2}\frac{Zm_e}{m_i} \lambda_B^2 \omega_p^2
    \end{equation}
    where $\beta_{b0}=v_{b0}/c$. The solution is exponential growth $\lambda_B(t) = \lambda_{B0} e^{\Gamma t}$ with a rate
    \begin{equation}\label{eq:gr1}
      \frac{\Gamma}{\omega_p} = \alpha \beta_{b0} \sqrt{\frac{Zm_e}{\delta m_i}}
    \end{equation}
    where $\delta = 4$ in the slab geometry \added{derived here}. A similar calculation for cylindrical geometry yields $\delta = 3$. Remarkably, for relativistic beams the growth rate is independent of their Lorentz factor $\gamma_{b0}$.

\begin{figure}
      \includegraphics[width=3.375in]{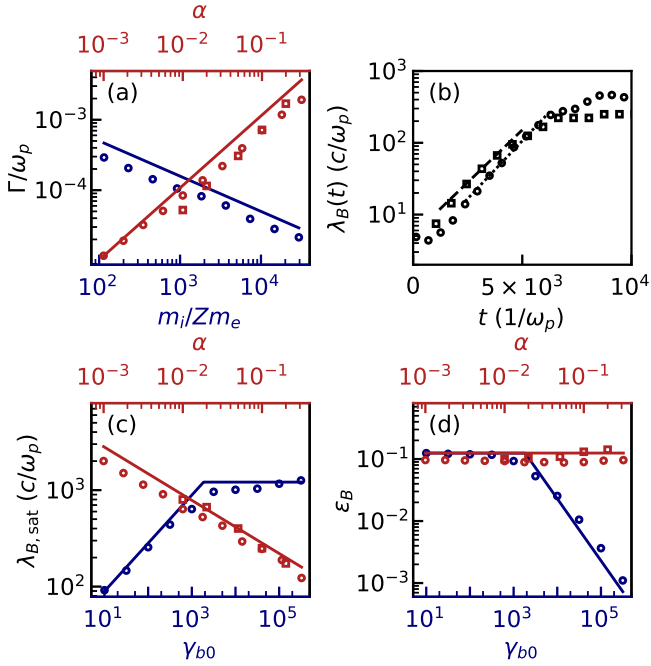}
      \caption{\label{fig:3}
        Comparison of analytic theory (solid lines) with 1D (circles) and 2D $y$-$z$ (squares) simulations for a dilute, ultrarelativistic electron beam propagating through an electron-ion background. \added{We use slab geometry theory given that the theoretical cylindrical growth rate differs only by $15\%$.} (a) Nonlinear instability growth rate. (b) Temporal evolution of the dominant magnetic field length scale $\lambda_B(t)$ for density ratio $\alpha = 0.1$, beam Lorentz factor $\gamma_{b0} = 1000$, and mass ratio $m_i/Zm_e = 1836$, with the growth rate taken from the magnetic energy overlaid as dotted (1D) and dashed (2D) lines. (c) Saturation magnetic field length scale $\lambda_{B,\text{sat}}$ compared to Eq. \ref{eq:sat_length}. (d) Saturation magnetization $\epsilon_B$ compared to Eq. \ref{eq:sat_eps}. All parameter scans in (a,c,d) use as fixed parameters $\alpha = 0.01$, $\gamma_{b0} = 1000$, and $m_i/Zm_e = 1836$.
        }
    \end{figure}

    \added{The theoretical growth rate agrees well with simulation results as indicated in Fig.~\hyperref[fig:3]{\ref*{fig:3}(a)}.
    We use purely transverse 1D and 2D ($y$-$z$) simulations to capture both slab and cylindrical geometries over a wide range of $\alpha$ and $m_i/Zm_e$ (including for reduced mass ratios $m_i/(Z m_e)<1836$ commonly used in previous numerical studies). In simulations that resolve the longitudinal ($x$) dimension similar agreement is observed [{\it e.g.} in Fig.~\hyperref[fig:1]{\ref*{fig:1}(e)} the measured growth rate is within $10 \%$ of the theoretical prediction]. In general, in purely transverse simulations the Weibel instability grows first and rapidly triggers the nonlinear instability, whereas when resolving the $x$-axis the oblique modes first heat the plasma, slowing the growth of the Weibel instability and delaying the onset of the nonlinear instability [see Fig.~\hyperref[fig:1]{\ref*{fig:1}(e)}]. The slower growth observed in some cases is primarily due to competition between cavities which lowers the pressure drop across the cavity wall; overall, the theoretical growth rate is within 40\% of the simulation results.}

    A key feature of this nonlinear instability is that the magnetic field strength and coherence length grow at the same rate. The temporal evolution of the dominant magnetic field length scale $\lambda_B(t)$ is shown for 1D and 2D $y$-$z$ simulations in Fig.~\hyperref[fig:3]{\ref*{fig:3}(b)} with the rate taken from the magnetic energy growth overlaid on top. The two rates match \added{closely during nonlinear instability growth starting at $t=1000~\omega_p^{-1}$}. We note that this $\lambda_B(t) \propto e^{\Gamma t}$ growth in length scale is very different from the $\lambda_B(t) \propto \sqrt{t}$ expected due to filament merging \cite{medvedev05}. 
    
    The nonlinear magnetic field growth will saturate when either the beam electrons or the background ions start to be significantly affected by the strong fields. We focus on the case of relativistic beams ($\gamma_{b0} \gg 1, \beta_{b0} \approx 1$), \added{which are} most interesting for high-energy astrophysical environments and laboratory plasmas associated with intense lasers and particle beams. In the limit where $\gamma_{b0} < m_i/Zm_e$, the beam electrons respond first; the saturation should occur when the beam electron gyroradius becomes comparable to the cavity radius $r_{gb} \approx \lambda_B/2$, also known as the Alfv\'en limit \cite{alfven39}. Considering that the average saturation magnetic field experienced by the particles in the cavity, $B_\mathrm{sat}$, is half of the peak field, we obtain a magnetic field spatial scale at saturation of $\lambda_{B,\mathrm{sat}} \approx \sqrt{8\gamma_{b0}/\alpha} (c/\omega_p)$, saturation magnetic field $B_\mathrm{sat} \approx \sqrt{\gamma_{b0} \alpha/2}(m_e \omega_p c/e)$, and associated magnetization $\epsilon_{B} \approx (1/2) B_\mathrm{sat}^2/(8\pi \gamma_{b0} n_{b0} m_e c^2) \approx 1/8$. The factor of $1/2$ in $\epsilon_B$ approximates the fact that roughly half of the volume is filled with cavities as seen in Fig.~\hyperref[fig:1]{\ref*{fig:1}(d)}.
    Notably, the magnetization is independent of both $\gamma_{b0}$ and $\alpha$, meaning that even for highly relativistic and very dilute beams a significant fraction of the beam energy is transferred to the magnetic fields; this is qualitatively different from previous results for the nonlinear Weibel instability, which reported saturated magnetic field levels of $\epsilon_B < 10^{-3}$ for $\alpha \leq 0.01$ \cite{sironi14,benedetti18}. 
    
    When $\gamma_{b0} > m_i/Zm_e$, the background ions respond to the presence of the strong magnetic fields before the beam electrons. The rapid growth of the magnetic field produces a strong inductive electric field. The ions in the cavity with density $n_i \approx \alpha n_{e0}$ accelerate due to this E-field and neutralize the beam current as they become relativistic. The E-field in the cavity can be estimated from Faraday's law as $E = \alpha \Gamma (\lambda_B/2)^2 (m_e \omega_p^2/e)$. By solving for the ion momentum $p_i$, we find that the E-field accelerates the ions to relativistic speeds, $p_i \approx m_i c$, at the point where the magnetic field reaches a saturation wavelength $\lambda_{B,\mathrm{sat}} \approx \sqrt{8(m_i/Zm_e)/\alpha} (c/\omega_p)$, amplitude $B_\mathrm{sat} \approx \sqrt{m_i/(2\alpha Zm_e)}(m_e \omega_p c/e)$, and magnetization $\epsilon_{B} \approx (1/8) \gamma_{b0} Zm_e/m_i$. 
    
    Together, these two mechanisms predict in the relativistic regime a saturation magnetic wavelength 
    \begin{equation}\label{eq:sat_length}
      \lambda_{B,\mathrm{sat}} \approx \sqrt{\frac{8}{\alpha} \textrm{min}\left\{\gamma_{b0},\frac{m_i}{Zm_e}\right\}} \frac{c}{\omega_p}
    \end{equation}
    and magnetization 
    \begin{equation}\label{eq:sat_eps}
      \epsilon_{B} \approx \frac{1}{8} \textrm{min}\left\{1,\frac{m_i}{\gamma_{b0} Zm_e}\right\}.
    \end{equation}
    where $\textrm{min}\{a,b\}$ is the smaller of $a$ and $b$. We can see that the two saturation mechanisms predict the same values when $\gamma_{b0} = m_i/Zm_e$. \added{Interestingly, when saturation is determined by the ion response, both $\epsilon_B$ and $\lambda_{B,\mathrm{sat}}$ increase with ion mass, in contrast to the growth rate}. 
    
    Fig.~\hyperref[fig:3]{\ref*{fig:3}(c,d)} \added{compares} our theoretical predictions for the saturation magnetic wavelength and magnetization with 1D and 2D \added{$x$-$y$} simulations over a wide range of $\alpha$ and $\gamma_{b0}$. The \added{results agree well with} Eqs. \ref{eq:sat_length} and \ref{eq:sat_eps} and clearly show the transition between the two saturation mechanisms.
    
\begin{figure}
      \includegraphics[width=3.375in]{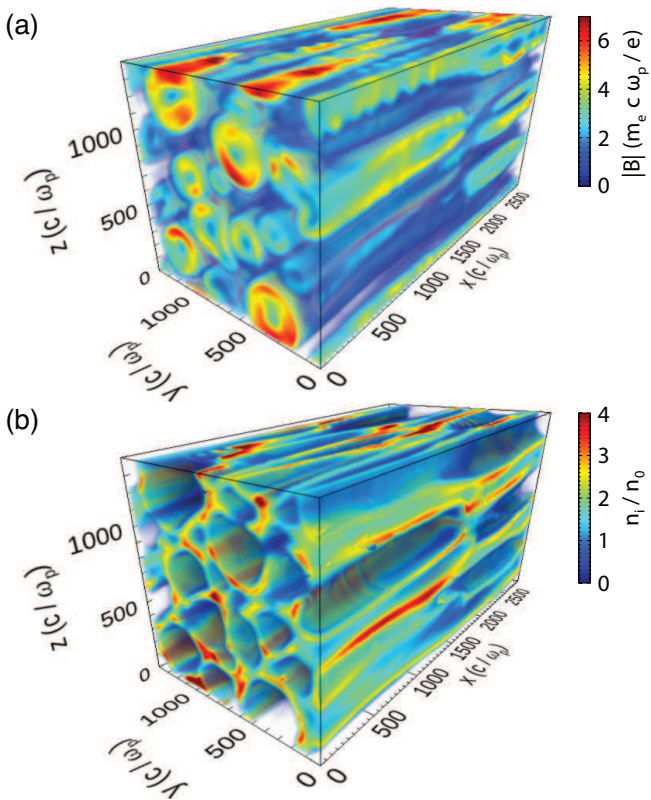}
      \caption{\label{fig:4}
        3D simulation results of a dilute, ultrarelativistic electron beam propagating (in the $x$-direction) through an electron-ion background. The magnetic field (a) and ion density (b) are reported at $t = 13000/\omega_p$ near saturation of the nonlinear instability, showing the amplification of large-scale magnetic fields and the generation of density cavities. The opacity scales linearly with the value.
      }
    \end{figure}

    This nonlinear instability is also robustly observed in 3D. Fig.~\ref{fig:4} shows simulation results for $\alpha = 0.05$, $\gamma_{b0} = 4000$, and $m_i/Zm_e = 1836$. The development of the instability proceeds similarly to that of Fig.~\ref{fig:1}. 
    The measured growth rate $\Gamma = 4.4\times 10^{-4}~\omega_p$, magnetization $\epsilon_{B} = 3.9$\%, and saturation magnetic field length scale $\lambda_{B,\textrm{sat}} = 400~c/\omega_p$, are in good agreement with the theoretical predictions of $\Gamma = 6.8\times 10^{-4}~\omega_p$, $\epsilon_{B} = 5.7$\%, and $\lambda_{B,\textrm{sat}} = 540~c/\omega_p$ and confirm that our analysis and the development of this new instability is robust in 3D systems.

    \added{The simulations presented consider initially uniform beam-plasma systems, which is a reasonable approximation for kinetic scales associated with astrophysical plasmas. We have performed additional simulations with finite-size beams (not shown here). This is motivated by the possibility of using picosecond kJ-class lasers to produce high-charge ($\gtrsim \murm {\rm C}$) electron beams \cite{albert20,shaw20}, which could enable the study of this instability in the laboratory. For example, for an electron beam with peak energy $\epsilon_b = 25$ MeV, energy spread $\Delta\epsilon_b/\epsilon_b = 1$, $5~\murm$C charge, $1$~ps duration, $50~\murm$m diameter, and 50 mrad divergence propagating in a hydrogen plasma with $n_{i0}=2\times10^{20}$ cm$^{-3}$ we observe that the nonlinear instability grows as predicted by theory and produces $100$~MG magnetic fields in the plasma after a 300 $\mu$m distance. A more detailed study of the applications of this instability to both laboratory and astrophysical plasmas is deferred to future work.}

    In conclusion, we have shown that a new electron streaming instability in relativistic dilute-beam systems can generate magnetic fields with orders of magnitude larger strength and spatial scale than previously reported. 
    This can have important implications for both astrophysical and laboratory scenarios. 
    \added{For example, in GRBs, it could mediate the amplification of magnetic fields upstream of the external relativistic shock to scales much larger than the plasma skin depth and help explain the observed synchrotron emission spectra \cite{medvedev99,gruzinov99,beloborodov05,ramirez-ruiz07}. It could also lead to enhanced particle scattering in blazar jets and potentially impact constraints on the intergalactic magnetic field \cite{broderick12,sironi14}.}
    \added{In the laboratory, this instability could} allow for more efficient transfer of electron beam energy to dense inertial fusion plasmas \cite{tabak94,silva02} and enable high energy compact radiation sources \cite{benedetti18}.
    

\begin{acknowledgments}
This work was supported by the U.S. Department of Energy SLAC Contract No. DEAC02-76SF00515, by the U.S. DOE Early Career Research Program under FWP 100331, by the DOE FES under FWP100182, and by the DOE NNSA Laboratory Residency Graduate Fellowship (LRGF) under grant DE-NA0003960. The authors acknowledge the OSIRIS Consortium, consisting of UCLA and IST (Portugal) for the use of the OSIRIS 4.0 framework and the visXD framework. Simulations were performed at Cori (NERSC) and Theta (ALCF) through ALCC computational grants.

\end{acknowledgments}


%

\end{document}